\documentstyle[11pt,epsfig]{article}
\title{Particle Creation from Vacuum by Lorentz Violation}
\author{Nima Khosravi\thanks{email: n-khosravi@sbu.ac.ir}\\
{\small {\it Department of Physics, Shahid Beheshti University, G.
C., Evin, Tehran 19839, Iran }}}

\begin{document}
\maketitle
\begin{abstract}
It is shown that the vacuum state in presence of Lorentz violation
can be followed by a particle-full universe that represents the
current status of the universe. In this model the modification in
dispersion relation (Lorentz violation) is picked up representing
the regime of quantum gravity. The result can be interpreted such
that the existence of the particles is an evidence for quantum
effects of gravity in the past. It is concluded that only the vacuum
state is sufficient to appear the matter fields spontaneously after
the process of semi-classical analysis. \vspace{5mm}\newline PACS:
04.60.-m (Quantum gravity), 04.62.+v (Quantum fields in curved
spacetime)
\end{abstract}
\maketitle
\section{Introduction}
The big-bang singularity is a prediction of classical general
relativity and it must be removed in the final theory. This
feature\footnote{And also black-holes.} causes concentration of the
community on the regimes that the general relativity (gravity) is
the dominant field with a very high density. From the other side our
knowledge of quantum physics says that in this regime due to
enormous density of a field the classical physics is broken and
quantum fields become essential. The big-bang and the black-hole are
theoretical evidences of quantum gravity, challenge the final theory
(at least its gravity sector). Until now different approaches to
quantize gravity have grown up e.g. string theory \cite{STbooks} and
loop quantum gravity \cite{QGbooks}. For such candidates or any
given new theory it is very crucial to be falsifiable i.e. it must
have some non-trivial accessible predictions, besides the
compatibility with their classical counterparts. However due to
complexity of quantum gravity it is so hard to calculate any
physical prediction comparing to the experiments. To make this
comparison possible the test theories are needed which fill the gap
between the full theories' predictions and the experiments. These
test theories should possess the structure of the main theories as
much as possible. Also there is (are) another layer(s) between the
full quantum gravity theories and test theories which is (are) named
falsifiable ``quantum gravity theory of not everything'' \cite{g}.
These theories make it possible to transit the great gap between
test theories and full theories step by step. As an example
noncommutative geometry \cite{con} can be mentioned.

As one of the candidate to model quantum gravity effects is a
deviation \cite{g,qglv,jac,nature,001,amelino} from the standard
dispersion relation among energy and momentum of a particle (Lorentz
violation) i.e. $\omega^2=k^2+m^2$ where $\omega$, $k$ and $m$ are
energy, momentum's norm and mass of the particle, respectively.
Actually, it is well-known that modification in dispersion relation
can appear as a consequence of discretization of space-time on a
lattice \cite{g,lattice}. Observationally in some cases the modified
dispersion relation's imprints can be found in cosmic ray spectrum
\cite{case}. This deviation can be represented by a modification in
the dispersion relation such that $\omega^2=k^2+m^2-\alpha^2 k^4$ as
a famous example \cite{jac}. However, this form of deformation is
not a unique choice e.g. introducing a cubic term has been studied
in details \cite{nature} and a more general form in \cite{001} which
contains an observer independent length beside velocity. The
coefficient $\alpha^2$ is a factor proportional to the minimum
length $\ell_P$ that makes the semi-classical limit to standard form
of dispersion relation trivial due to exiting from quantum gravity
regime by taking $\ell_P \rightarrow 0$. It is worth mentioning that
this prediction is a candidate to go further in the phenomenology of
quantum gravity \cite{g,amelino}. As mentioned in the above, the
phenomenological aspects of quantum gravity is a controversial issue
in theoretical physics. Though it is arguable to suppose that the
regime of quantum gravity does not make any senses to do some
repeatable experiments by humankind but it is believable to accept
domination of quantum gravity in early universe. Therefore
illustration of this regime may shed some lights on the nature of
quantum gravity.

In general to achieve properties of a physical system, knowledge of
the dynamical rules besides the correct initial conditions, are
needed. For our universe the common believing is that the initial
state is a quantum gravitational state due to domination of
gravitational field in the very early times. This state has reached
to the present state such that the universe contains the particles
and also in lack of the quantum effects of gravity\footnote{To be
more precise, in very very tiny effects of quantum gravity.}. So the
question is that how the particle-full universe has risen up from an
unknown initial quantum gravitational state? Note that the initial
state is an unknown state at least at the present. But as it is
usual, with the similar predictions the simpler choice is the better
choice. At least in lack of any knowledge about the correct initial
state, the simpler choice makes the theory calculable resulting in a
general understanding of the theory however it is not complete. The
simplest choice for the initial state is the vacuum state, if it
does not fall in a trivial prediction. If this simplest choice can
predict non-trivial particle-full present status of the universe
then the result is very interesting and remarkable. Note that the
vacuum state is not only the simplest choice but also is a special
choice. This choice makes the proposal of creation from nothing
meaningful \cite{nothing}. The idea of vacuum creation theory has
been considered also in different aspects such that under the action
of strong fields it is possible to create particle-anti particle
plasma system \cite{030,031}. As mentioned in \cite{031} the time
dependent masses can cause particle creation that makes these kind
of models comparable to our model which has a time dependent
dispersion relation.

In this work, in a toy model we have shown that the above argument
can be obtainable. To construct our toy model we pick up the method
of particle creation in the context of the quantum field theory in
curved space-times \cite{birreldavis}. The deformed dispersion
relation is used to model the quantum gravitational field i.e.
$\omega^2=k^2+m^2-\alpha^2 k^4$. We will propose that the quantum
gravity parameter $\alpha^2$ has a time dependence such that for
very early times be one and vanishing for late times. This evolution
behavior is appropriate to study the effects of very early quantum
gravity in present time. But it should be mentioned that the
function of $\alpha(t)^2$ is taken arbitrarily, in lack of any full
quantum gravity, such that the equations become solvable
analytically.

\section{Model}
It is generally believed that the notion of particle and as a
consequence the notion of vacuum in quantum field theory is not a
straightforward manner specially in curved space-times. The crucial
part of the definition of states in field theory is the selection of
the basis. A field can be expanded due to the appropriate basis
$u_k(x)$ as follow
\begin{eqnarray}\label{field}
\varphi(x)=\sum_k \left(a_k u_k(x)+a_k^*u^*_k(x)\right)
\end{eqnarray}
where $a_k$ is a complex number, $x$ and $k$ are four-vectors of
position and momentum respectively. In the Minkowskian space-time
the choice is trivially $e^{\pm i \vec{k}. \vec{x}}e^{\pm i \omega
t}$ such that $\omega^2={\vec{k}}^2+m^2$. The next step is the
quantization that is being constructed by transition from complex
number coefficients $a_k$ to their corresponding operators
$\hat{a}_k$ and consequently $a_k^*$ to $\hat{a}^{\dagger}_k$, such
that the commutation relation $[\hat{a}_k,\hat{a}_{k'}^{\dagger}]=i
\hbar \delta_{ij}$ is satisfied by $\hat{a}_k$ and
$\hat{a}^{\dagger}_k$. Due to this relation $\hat{a}_k$ and
$\hat{a}^{\dagger}_k$ can be interpreted as annihilation and
creation operators. Finally an $n_k$-particle state with momentum
$k$, $|n_k>$, is being defined by
$\hat{N}_k=\hat{a}^{\dagger}_k\hat{a}_k|n_k>=n_k|n_k>$. In the above
we reviewed very quickly the structure of definition of states
containing particles. As mentioned before the starting point is
definition of the appropriate basis that is not trivial for curved
space-times \cite{birreldavis}. In general cases the symmetries can
help in defining the useful basis. This rapid review was needed
entering to our model.

In our model we will suppose a Minkowskian space-time as the
background but the dispersion relation is different for two sides of
the time interval. For the very early times, i.e. for $t\rightarrow
- \infty$, it has the form $\omega^2=\vec{k}^2+m^2-\alpha_0^2
\vec{k}^4$ where $\alpha_0^2$ is the Lorentz violation parameter
represents the quantum gravity regime in our model. And for very
late times, i.e. $t\rightarrow + \infty$, the dispersion relation
becomes the standard one, $\omega^2=\vec{k}^2+m^2$. For the late
times the natural choice of the basis is $e^{\pm i \vec{k}.
\vec{x}}e^{\pm i \omega t}$ where $\omega^2=\vec{k}^2+m^2$. But for
the early times these basis are not the suitable ones since the
dispersion relation $\omega^2=\vec{k}^2+m^2$ has not any
significance in this time region. The natural alternative for this
region of time is $e^{\pm i \vec{k}. \vec{x}}e^{\pm i \omega t}$
where $\omega^2=\vec{k}^2+m^2-\alpha_0^2 \vec{k}^4$. Since the
definitions of basis for these two different regions are not
equivalent then the equivalence of the vacuum's notion for them is
not a trivial concept and must restudy again. This feature can cause
a deviation from the initial vacuum state and the final vacuum
state. This deviation can be interpreted as particle creation during
transition from initial to final state. Note that this kind of
interpretation is a standard one in quantum field theory in curved
space-times \cite{birreldavis}. The aim of this paper is to study
this concept. To do more, supposing the time evolution of the
Lorentz violation parameter is
\begin{eqnarray}\label{alpha}
\alpha(t)^2=\frac{\alpha_0^2}{1+e^{t}},
\end{eqnarray}
where $\alpha_0^2$ is the initial value of the Lorentz violation
parameter and the general behavior is such that the parameter for
the early times and the late times satisfies our above propositions.
Otherwise, the form of the function is picked only to make the
equations exactly solvable. It must be noted that the origin of this
form of time dependence has not been discussed in this work and it
seems that this subject belongs to the quantum gravity and specially
that branch discussing on the semi-classical limit of quantum
gravity\footnote{It is not too bad mentioning this branch of
research is an active part without any final conclusion
\cite{QGbooks,rovelli}.}. We consider a mass-less scalar field
without any loss of generality. To reach to the Klein-Gordon
equation it is sufficient to replace $\omega$ and $\vec{k}$ in the
dispersion relation by their operator form $-i\partial_t$ and
$-i\partial_{\vec{x}}$ respectively. So the Klein-Gordon
equation\footnote{It is worth mentioning that to reach to
Klein-Gordon equation the quantization process is crucial. The
details of reaching to above equation is in \cite{ehsan}. It should
be notice that in the quantization process the concept of symmetry
transformations in presence of Lorentz violation are not trivial.
This subject is considered in details in \cite{reza}.} becomes
\begin{eqnarray}\label{KGeq}
\left[\partial^2_t-\partial^2_{\vec{x}}-\alpha(t)^2\partial^4_{\vec{x}}\right]\varphi(x)=0,
\end{eqnarray}
where $x$ is position four-vector. Letting $\varphi(x)\propto e^{ i
\vec{k}. \vec{x}}T_k(t)$ reduces the above equation to
\begin{eqnarray}\label{reducedKGeq}
\left[\partial^2_t+k^2-\alpha(t)^2k^4\right]T_k(t)=0,
\end{eqnarray}
where $k=|\vec{k}|$. The solution of the above second order
differential equation is
\begin{eqnarray}\label{solutions}
T_k(t)=C_1 e^{-i
\left(\sqrt{k^2-\alpha_0^2k^4}\right)t}{_2F_1}\left(a,b;c;-e^t\right)
+C_2 e^{+i
\left(\sqrt{k^2-\alpha_0^2k^4}\right)t}{_2F_1}\left(b^*,a^*;c^*;-e^t\right)
\end{eqnarray}
where $_2F_1$ is the hypergeometric function and
\begin{eqnarray}\label{parameters}
a&=& -i k-i\sqrt{k^2-\alpha_0^2k^4}=-i(\omega_{in}+\omega_{out})\nonumber\\
b&=& +i k-i\sqrt{k^2-\alpha_0^2k^4}=-i(\omega_{in}-\omega_{out})\\
c&=&1-2
i\sqrt{k^2-\alpha_0^2k^4}=1-i\hspace{.5mm}2\omega_{in}\nonumber
\end{eqnarray}
To go further we concentrate to the behavior of the both infinite
limits. At the first for the very early times, $t \rightarrow
-\infty$, the above solution reduces to
\begin{eqnarray}\label{-solutions}
T_k(t\rightarrow -\infty)=C_1 e^{-i \omega_{in}t}+C_2 e^{+i
\omega_{in}t},
\end{eqnarray}
where the identity $_2F_1(a,b;c;0)=1$ for arbitrary $a$, $b$ and $c$
is used \cite{abr}. The result is fully in agreement with our
expectation since for the very early times the energy is
$\omega_{in}=\sqrt{\vec{k}^2-\alpha_0^2 \vec{k}^4}$. So the first
term in (\ref{-solutions}) can be interpreted as the temporal part
of $u_k^{in}$ that we will need it in the following calculations.
For the second infinite limit, the very late times, the solution
becomes more complicated such that
\begin{figure}[th]
\centerline{\includegraphics[width=10cm]{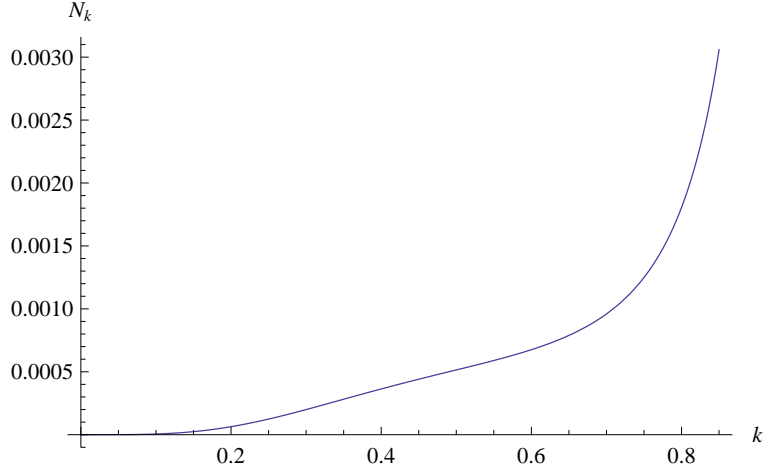}}
\caption{\label{fig1}\footnotesize The parameter $\alpha_0^2=1$ is
positive then the plot is not shown for the forbidden region where
the $\omega_{in}$ is imaginary (in this case $k>1$ represents
forbidden region).}
\end{figure}\begin{eqnarray}\label{+solutions}
T_k(t\rightarrow +\infty)&=&\left[C_1
\frac{\Gamma(c)\Gamma(a-b)}{\Gamma(a)\Gamma(c-b)}+ C_2
\frac{\Gamma(c^*)\Gamma(b^*-a^*)}{\Gamma(b^*)\Gamma(c^*-a^*)}\right]\times e^{-i \omega_{out}\hspace{.5mm} t}\nonumber\\
&+&\left[C_1 \frac{\Gamma(c)\Gamma(b-a)}{\Gamma(b)\Gamma(c-a)}+C_2
\frac{\Gamma(c^*)\Gamma(a^*-b^*)}{\Gamma(a^*)\Gamma(c^*-b^*)}\right]\times
e^{+i \omega_{out}\hspace{.5mm} t},
\end{eqnarray}
where $\Gamma(x)$ is the gamma function. The important point in the
above solutions is that for incoming waves $e^{-i
\omega_{out}\hspace{.5mm} t}$ there is a combination of the two
terms of the general solutions (\ref{solutions}) i.e. $C_1$ and
$C_2$ both appear in the coefficient of the incoming waves and the
same is true for the outgoing waves. This means that the vacuum
state of the very early universe, $t \rightarrow -\infty$, does not
coincide to the vacuum state of the very late times, $t \rightarrow
+\infty$. As mentioned before, this interpretation is a standard
interpretation in the subject of quantum field theory in curved
space-times \cite{birreldavis}. For continuing we must calculate the
Bogolubov coefficients, $\gamma$ and $\beta$, as the following
\cite{birreldavis}
\begin{eqnarray}\label{bog}
u_k^{in}(x)=\gamma_ku_k^{out}(x)+\beta_ku_{-k}^{out*}(x),
\end{eqnarray}
where $x$ is the position four-vector. Note that $u_k^{in}$ and
$u_k^{out}$ are the first terms in relation (\ref{field})
corresponding to annihilation operator for $in$-region and
$out$-region respectively. The non-vanishing $\beta$ results in
incoincidence vacua for $in$-region and $out$-region e.g. in our
model, the very early times and the very late times respectively. In
mathematical language due to relation (\ref{-solutions}) in our
model $u_k^{in}(x)$ is the first term in relation (\ref{solutions})
times $e^{ i \vec{k}. \vec{x}}$ i.e. it contains only the $C_1$
factor. But for the $out$-region the result is more complicated such
that due to relation (\ref{+solutions}), $u_k^{out}(x)$ equals to a
combination of both of terms in (\ref{solutions}). The temporal part
of the solutions for both of the $in$- and $out$-regions are
\begin{eqnarray}\label{-+}
T_k^{in}(t)&=&\frac{1}{(4 \pi
\omega_{in})^{\frac{1}{2}}}\hspace{2mm} e^{-i
\omega_{in}t} \hspace{1mm}{_2F_1}\left(a,b;c;-e^t\right)\nonumber\\
T_k^{out}(t)&=&\frac{(4 \pi \omega_{out})^{\frac{1}{2}}}{(4 \pi
\omega_{in})}\times\\
&&\left(e^{-i \omega_{in}t}
\frac{\Gamma(c^*)\Gamma(a^*-b^*)}{\Gamma(a^*)\Gamma(c^*-b^*)}\hspace{1mm}{_2F_1}(a,b;c;-e^t)-e^{+i
\omega_{in}t}
\frac{\Gamma(c)\Gamma(b-a)}{\Gamma(b)\Gamma(c-a)}\hspace{1mm}{_2F_1}(b^*,a^*;c^*;-e^t)\right)\nonumber,
\end{eqnarray}
such that by taking the limits, the above solutions reduce to the
first terms in their counterpart relations (\ref{-solutions}) and
(\ref{+solutions})\footnote{It is worth mentioning again that for
the second relation, $T_k^{out}$, the combination has picked up such
that the above relation reduces to its asymptotic counterpart in
(\ref{+solutions}).}. Note that in the above results the prefactors
guaranties the normalization of the basis. The second Bogolubov
coefficient contributes to show the spectrum of the created
particles with respect to the energy i.e. the $|\beta_k^2|$ is the
number of particles with energy $k$, is in the following form
\begin{eqnarray}\label{number}
N_k=\beta_k^{2}&=&\frac{\sinh^2(\pi
(\omega_{out}-\omega_{in}))}{\sinh(2 \pi \omega_{in})\times \sinh(2
\pi \omega_{out})}.
\end{eqnarray}
The above result can be deduced by some algebra (see the Appendix)
from definition of Bogolubov coefficients (\ref{bog}) that has been
used for relations (\ref{-+}). The figures show the above number
density spectrum for different values of $\alpha_0^2$. Now let us
examine the behavior of the result in well known limits. For
$\alpha_0^2=0$ the expectation is, vanished $N_k$ due to no
difference between the early and late times. Obviously since in this
case $\omega_{in}=\omega_{out}$ it causes vanishing
``$\sinh(\pi(\omega_{in}-\omega_{out}))$" in the numerator of the
fraction results in vanished $N_k$, as expected. Another point is
that we have two different choices for $\alpha_0^2$, a positive one
and a negative one. It must be noted that for the positive one, we
must restrict the plots to an upper-bound for $k$'s since for the
greater values of $k$, $\omega_{in}$ becomes an imaginary number
that makes no sense in our conclusions. But for the negative values
of  $\alpha_0^2$ the spectrum is valid for all the $k$'s. The figure
1 shows the behavior of the $N_k$ with respect to $k$ for a positive
$\alpha^2_0$. In this case as mentioned before, there is a forbidden
region due to non-real values of energy $\omega_{in}$. The second
figure presents the spectrum of the number of created particles with
respect to their energy but for two negative values of $\alpha^2_0$.
It is obvious from the plots that in this case, the maximum of the
spectrum is changed by different values of Lorentz violation
parameters. The energy, that has the maximum number of created
particle, decreasing due to increasing of the absolute value of
$\alpha_0^2$. In this case for smaller value of $\alpha_0^2$ the
plot falls and for $\alpha_0^2\rightarrow 0$ it coincides to $N_k=0$
for all $k$'s, as one expected.
\begin{figure}[th]
\centerline{\includegraphics[width=10cm]{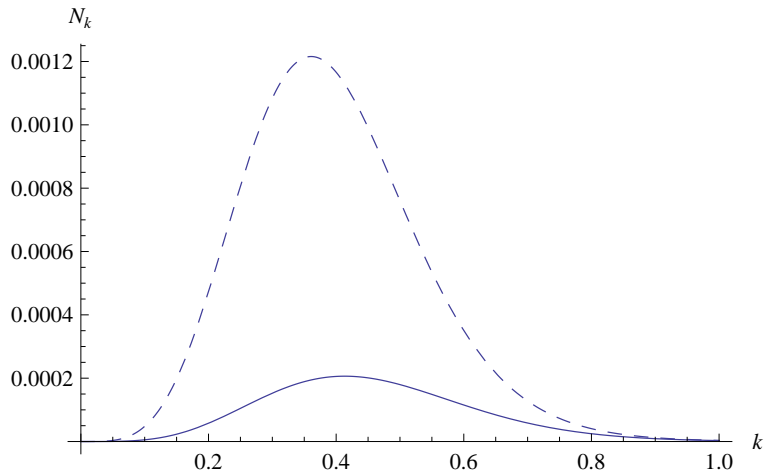}}
\caption{\label{fig1}\footnotesize This plot is for two negative
$\alpha^2_0$ with no forbidden region. The solid line presents the
spectrum of created particles for $\alpha_0^2=-1$ and the dashed
line for $\alpha_0^2=-3$.}
\end{figure}
\section{Conclusions}
We have studied a toy model to describe the effects of Lorentz
violation in particle creation subject \cite{ehsan} in presence of a
time dependent deformed dispersion relation. In this paper, we have
shown that the existence of Lorentz violation in very early times'
vacuum can result in the existence of particles but in the absence
of any Lorentz violation. In the other words we have combined two
legitimate individual ingredients, modification of the dispersion
relation \cite{jac,nature,001} and time variation of a fundamental
parameter \cite{bd}, to peruse any interesting features. According
to above discussions and close relation between Lorentz violation
and quantum gravity a suggestion can be introduced that: quantum
gravitational vacuum causes a particle-full state in classical
gravity\footnote{This suggestion is not provable until reaching a
full theory of quantum gravity. But the current results, in context
of a toy model, make possible the validity of this claim.}. This
final state is obtainable by making a semi-classical limitation
process. These particle-full states are obtained by taking
semi-classical limits of quantum gravitational vacuum. This means
particles are only some excitations of quantum gravitational vacuum,
that is not an obsolete idea in theoretical physics \cite{briad}. In
this viewpoint, the particles are the evidences of the past
existence of Lorentz violation (or, quantum structure of the
geometry (gravity)). The above argument is totally in agreement with
some kinds of interpretation of wave function in quantum cosmology
such that ``probabilistic evolutionary process" interpretation
\cite{pep}. Since our model is a toy model and far from complete
quantum gravity theory, then to see the correctness of our proposal
in this paper we need waiting until full quantum gravity theory.
This can also shed lights on the origin of the time-dependence of
Lorentz violation parameter that is an ambiguity in our model.
Finally, it is again worth to mention that in our model only the
Lorentz violation is picked up presenting the quantum gravity
regime\footnote{As mentioned before, this choice makes questionable
the validity of the concluding remarks in all the quantum gravity
regime. But at least, it sheds some lights on the behavior of this
unaccessible regime by analytical results.}. To discuss more
accurate we need to study all the quantum gravitational effects in
the presence of the curved background and its dynamics.

At the end it is proper to consider that the origin of the present
particles and even the large scale structures can be explained by
proposing an inflationary era in the early universe
\cite{inflation}. Maybe it causes to think that the inflationary
scenario makes the results of this paper irrelevant. But it is
important to note that the inflationary era is not the first stage
after the big-bang where quantum gravitational effects are dominant
\cite{gas}. Our conclusion is that the created particles by Lorentz
violation are just before the inflationary era i.e. in contrast to
the standard approach in inflationary models the initial state for
this era is not vacuum state. In this viewpoint \cite{gas} the
natural question will be how one can find any traces of these
particles after inflationary era in the last scattering surface or
even at present time? It seems that at least in our model since the
old\footnote{Before inflation.} particles are created by a geometric
effect then they show themselves by a geometric feature e.g.
gravitational waves \cite{camelia}. This way of thinking is still
open and in this work we did not consider this problem e.g. the
effects on CMB temperature fluctuations and etc.

\vspace{10mm}\noindent\\
{\bf Acknowledgments}\vspace{2mm}\noindent\\ I would like to thank
H. Firouzjahi, S. Jalalzadeh and M. M. Sheikh-Jabbari for
discussions and also anonymous referee for useful comments. This
work was supported in part by the Bonyad-e-Nokhbegan grant.
\\\\\appendix{\textbf{Appendix}}

In the body of the paper we have used alternatively the following
identities among the hypergeometric functions
\begin{eqnarray}\label{identity1}
F(a,b;c;z)=F(b,a;c;z)&=&\frac{\Gamma(c)\Gamma(b-a)}{\Gamma(b)\Gamma(c-a)}(-z)^{-a}F(a,1-c+a;1-b+a;\frac{1}{z})\nonumber\\
&+&\frac{\Gamma(c)\Gamma(a-b)}{\Gamma(a)\Gamma(c-b)}(-z)^{-b}F(b,1-c+b;1-a+b;\frac{1}{z})\nonumber
\end{eqnarray}
and the following properties for gamma functions
\begin{eqnarray}\label{identity2}
\Gamma(1+i y)&=&iy\Gamma(i y)\nonumber\\
\Gamma(i y)\Gamma(-i y)&=& \frac{\pi}{y \sinh(\pi y)}\nonumber
\end{eqnarray}
where $y$ is a real number.

\end{document}